\documentclass[12pt]{iopart}
\usepackage{graphicx}

\usepackage{color}
\usepackage[final]{hyperref}
\usepackage{amssymb}
\usepackage{textcomp}

\begin{document}


\title[Sclafani et al.]{Sensitivity of a superconducting nanowire detector for single ions at low energy}
\author{Michele Sclafani, Markus Marksteiner, Fraser McLennan Keir}
\address{Vienna Center for Quantum Science and Technology, Faculty of Physics, University of Vienna, Boltzmanngasse 5,
A-1090 Vienna, Austria}
\author{Alexander Divochiy´, Alexander Korneev, Alexander Semenov, Gregory Gol'tsman}
\address{Department of Physics, Moscow State Pedagogical
University, M. Pirogovskaya Street 1, Moscow 119992, Russia}
\author{Markus Arndt}

\address{Vienna Center for Quantum Science and Technology, Faculty of Physics, University of Vienna, Boltzmanngasse 5, A-1090 Vienna, Austria. 

email: markus.arndt@univie.ac.at, 

homepage: http://www.quantumnano.at}

\begin{abstract}
We report on the characterization of a superconducting nanowire detector for ions at low kinetic energies. We measure the absolute single particle detection efficiency $\eta$ and  trace its increase with energy up to $\eta = 100$\,\%. We discuss the influence of noble gas adsorbates on the cryogenic surface and analyze their relevance for the detection of slow massive particles. We apply a recent model for the hot spot formation to the incidence of atomic ions at energies between 0.2-1\,keV. We suggest how the differences observed for photons and atoms or molecules can be related to the surface condition of the detector and we propose that the restoration of proper surface conditions may open a new avenue to SSPD-based optical spectroscopy on molecules and nanoparticles.
\end{abstract}

\maketitle

\section{Introduction}
 Over the last decades, several different classes of cryogenic devices have been proposed for the detection of ionized \cite{Twerenbold1996b, Frank1999, Suzuki2010} or even neutral \cite{Marksteiner2009} particle beams. This comprises cryogenic bolometers \cite{Cavallini1967}, superconducting tunneling junction devices (STJ) \cite{Shiki2009a, Ohkubo2011} as well as normal metal insulator superconductor junctions (NIS) \cite{Hilton1998}.

 Here, we focus on one specific variant which relies on the fact that the impact of energetic particles may create localized resistive regions in a
 superconducting nanowire; which can be visualized as nanosecond voltage pulses when the wire is driven by a constant current source.  These detectors were originally designed for telecommunication purposes as single-photon sensors in the infrared radiation band. The literature therefore usually refers to them as \emph{superconducting single photon detectors} (SSPD) \cite{Goltsman2001}.

 Since any device that is highly responsive to small energy changes may also be considered as a detector for massive particles, several research groups have focused on the application of SSPDs in the context of mass spectroscopy \cite{Marksteiner2009,KojiSuzuki2008, Casaburi2009}. In many experiments clusters or molecules are generated as isolated ionized particles and accelerated to energies in the range of typically 10-30\,keV. The time-of-flight in such devices is then characteristic for the particle mass and the only requirement for the final detector is to record the object impact with high efficiency and a time resolution on the nanosecond scale. The general feasibility of this concept has already been demonstrated before and the interest has recently been focused on the exploration of detector characteristics such as optimal size, materials, geometry and  sensitivity \cite{Zen2009, A.Casaburi2011}.

 While the SSPD detection efficiency for electromagnetic radiation was reported to vary between $\eta = 10^{-4}..10^{-1}$ \cite{Goltsman2009} it has often been argued that
 the impact of massive particles should be detectable with certainty \cite{Ohkubo2008} once the kinetic energy suffices to heat the nanowire above the critical temperature.
 In many practical cases this assumption is easily fulfilled with particle energies exceeding several keV.
 In our present contribution we show, however, that this assumption may fail already for particle energies below a few hundred electronvolts:
 In many practical cases an atom or molecule arriving at the detector surface will not transfer all of its kinetic energy to the electronic system of the superconducting nanostripe. The incident particle may rather bounce off the surface or convert kinetic energy into cage deformation and internal heat during its impact.
 Massive particles will also interact with the nanowire phonon system directly in addition to electronic excitations, as usually discussed in SSPD photodetection.

 Our present work focuses on the role of surface adsorbates which have remained neglected in all previous studies, so far. They prove to be crucial as they strongly attenuate the energy transfer from the impacting particles to the SSPD cooper pairs. Liquefied or frozen gases on top of the nanowire are expected to be irrelevant for the detection of photons but they open new dissipation channels for low energy nanoparticles.
 In the following we will show that the SSPD design can nevertheless be highly sensitive to low energy ions.

\section{Experimental setup}

\begin{figure} [bh]
  \centering
 \includegraphics[width=8.5cm]{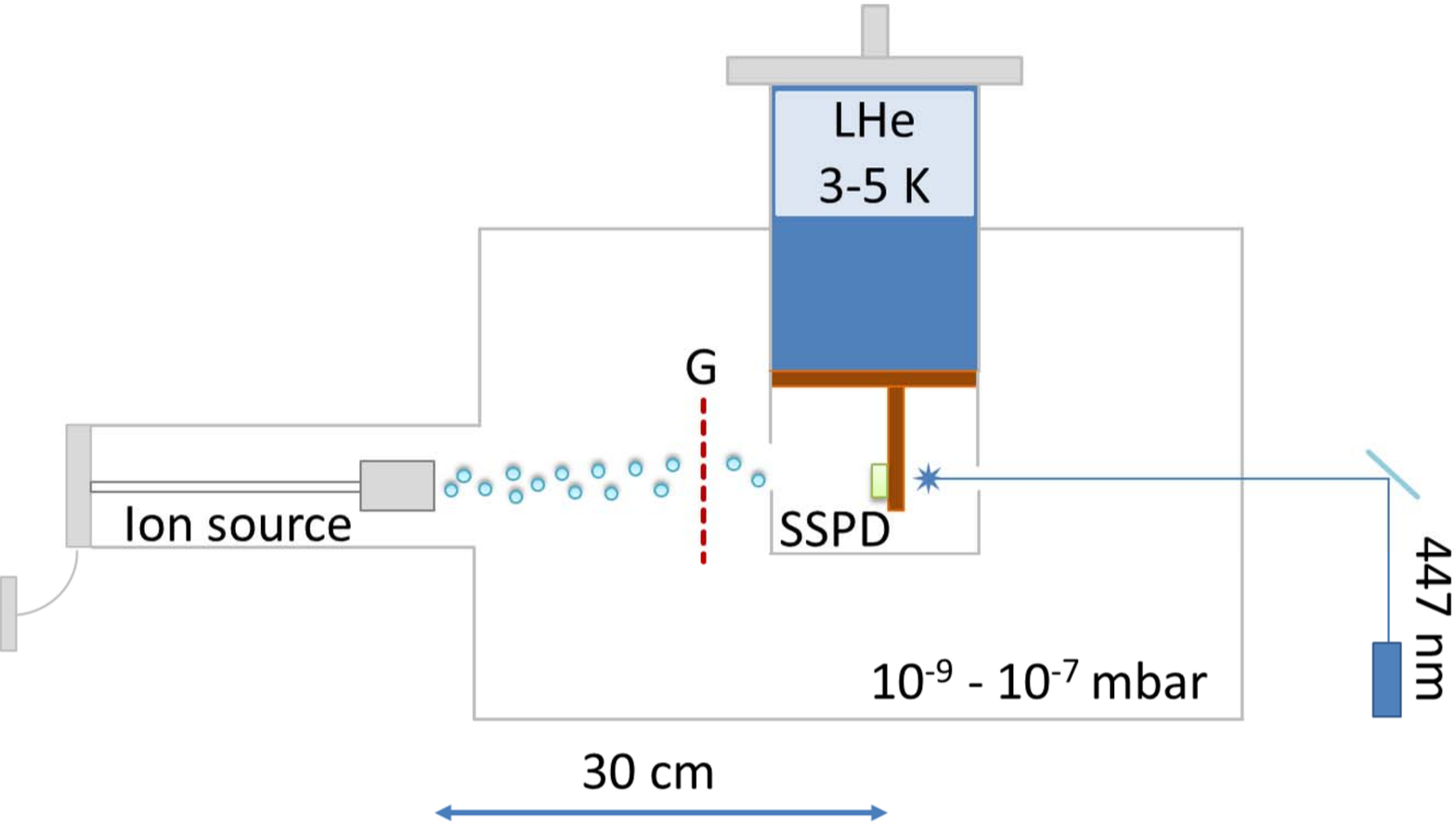}
  \caption{Setup for low energy ion detection using a superconducting nanowire detector.
  The ion source delivers noble gas atoms, here in the energy range between 200 and 1000\,eV. The ions can be blocked by applying a voltage to the wire mesh $G$.  The SSPD chip is mounted on a stage in thermal contact with a liquid helium cryostat whose temperature can be varied between 3-5\,K by pumping on the liquid helium bath. The system in kept under high vacuum. A Faraday cup can be installed in place of the SSPD for calibration measurements. The blue diode laser is used to desorb gases for them chip by heating it from the back. }
 \label{Fig1}
\end{figure}

  The experimental setup is schematically illustrated in Figure \ref{Fig1}: a clean beam of atomic ions is generated by a sputter gun (IQE 11/35), whose extraction energy is varied between 0.2 and 1\,keV. The ion current can be controlled via the electron current in the ionizer stage. Our experiments were performed with a number of different noble gases and also SF$_6$. In the following, we limit our presentation to the case of He$^{+}$, first because the same qualitative behavior was observed for all noble gases and second since helium adsorbates are removed more easily from the SSPD surface, if needed.
  The detector is placed approximately 30\,cm downstream of the ion gun and it is mounted on a copper holder in good thermal contact with a liquid helium cryostat. Our nanowire detector consists of a 4\,nm thick and 100\,nm wide NbN film that was sputtered in a long meander onto a sapphire substrate. The detector covers an area of 20$\times$20 $\mu$m$^{2}$ with a filling factor of f $\simeq$ 50\,$\%$. We operate the device in high vacuum (10$^{-7}$-10$^{-9}$\,mbar) and the detector temperature can be varied between T=3-5\,K, i.e. well below its critical temperature of T$_{c}$ = 10\,K. In order to prevent thermal fluctuations and to block photons from outside, the detector is shielded by an aluminum cylinder covered with mylar foil.

  A wire mesh (\emph{G} in Figure \ref{Fig1}) can be optionally inserted in front of the detector entrance window to discriminate between photons and particles. We verified that the detector signal goes back to the background value when the mesh potential equals the ion extraction voltage. This proves that ions were recorded rather than neutral atoms or even photons.  The SSPD can be optionally replaced by a Faraday cup (KimballS66) connected to an ammeter whose signal provides a reliable particle number calibration at elevated ion currents.

\section{Hot-spot model}

Throughout recent years, different models have been suggested to explain the underlying detection process \cite{Semenov2001,Semenov2005, Maingault2010}. The 'hot-spot' model was found to be in good qualitative agreement with experimental observations both for photons and for ions \cite{Verevkin2002,Suzuki2011}: when a molecule hits the superconducting film an energy transfer occurs. The released energy leads to the break of a Cooper pair and therefore to the formation of a high energy electron and of an electron whose energy is close to the superconducting gap $\Delta$. The high energy electron releases its energy producing in turn high energetic electrons and phonons which effectively break other Cooper pairs. In this avalanche process the concentration of quasiparticles increases on the picosecond time scale and suppresses the superconductivity in a local region, the so called "hot-spot". The dimension of this hot-spot grows as the quasi-particles diffuse into their surroundings. The supercurrent is expelled from the hot-spot and flows around its outside. As soon as the local current density exceeds its critical value \textit{j}$_{c}$, an ohmic resistance is established over the entire wire cross-section. This manifests itself as a measurable voltage peak across the detector.

\begin{figure} [h]
  \centering
 \includegraphics[width=8.5cm]{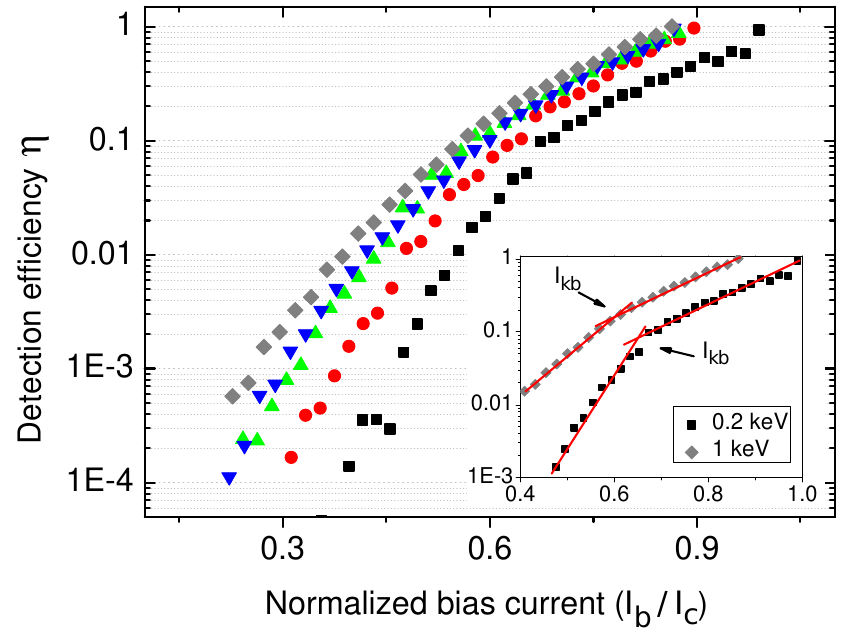}
  \caption{Absolute detection efficiency as a function of the normalized bias current (I$_{b}$/I$_{c}$) for He$^{+}$ at $E_{\mathrm{kin}}$=200\,eV (square), 400\,eV (circle), 600\,eV (up-triangle), 800\,eV (down-triangle) and 1000\,eV (diamond). \textbf{inset}: zoom into the minimum and maximum energy curve: in agreement with \cite{Verevkin2002} we define the slope point \emph{I$_{kb}$} for estimating the hot spot radius. Lines and arrows are plotted to guide the eye.}
 \label{Fig2}
\end{figure}

An experimental verification of the hot spot model is usually based on the observation of the detection efficiency $\eta$ as a function of the detector bias current $I_b$. Figure~\ref{Fig2} traces $\eta$ for five He$^+$ impact energies below 1\,keV. We note that we measure the \emph{absolute detection efficiency}, defined by the ratio of the SSPD ion counts $S_\mathrm{SSPD}$ and the number of ions seen by the optional calibrating Faraday cup $S_\mathrm{FC}$ according to $\eta_{abs}= (S_\mathrm{SSPD}-S_\mathrm{dark})/f S_{FC}$.  In this relation we also corrected for dark counts $S_\mathrm{dark}$ and the meander filling factor $f$.
The dark counts could be related to unshielded stray photons, thermal effects,electrical noise or even due to the dissipation of vortices \cite{Berdiyorov2009,Bulaevskii2011}.

A number of observations can be made in comparison to earlier work \cite{Suzuki2011}: First, we confirm the hypothesis that narrower nanowires are more energy sensitive than wider structures. While the former experiments detected no signal for ion energies lower than $E_{\mathrm{kin}}=600\,$eV in a detector with w=800\,nm wide stripes, we still see a measurable signal at 200\,eV for wires with w=100\,nm.  In contrast to the earlier work, we find no threshold behavior either. This gives rise to the hypothesis that more sensitive electronics will also allow us to see even slower particles.
In the present experiments the minimum test energy was determined by ion source.
At high energies and at high bias currents the detector saturates with an absolute detection efficiency of 100\%, as expected.

Qualitatively one may identify two regions of approximately exponential increase in $\eta(I_b)$, approximated by straight lines in the logarithmic plot, which are joined around a 'kink bias current' $I_{kb}$, as shown in the inset of Figure~\ref{Fig2}.
At fixed I$_{b}$/I$_{c}$, the increase of $\eta$ with $E_{\mathrm{kin}}$ is attributed to a growth of the normal-conducting hot-spot. When we fix the energy, the detection efficiency increases with the bias current, in agreement with Figure~\ref{Fig2}.

In contrast to the procedure chosen by Suzuki et al.~\cite{Suzuki2011}, we follow Verevkin et al.~\cite{Verevkin2002} and derive an estimate for the hot-spot radius \emph{R} from the ratio between the kink bias current \emph{I$_{kb}$} and the critical current \emph{I$_{c}$}:
\begin{equation}\label{equ1}
R=\frac{w}{2} \left(1-\frac{I_{kb}}{I_{c}} \right)
\end{equation}

From curves as shown in the inset of Figure~\ref{Fig2} we estimate the hot-spot radius R$_{\mathrm{ion}}$ for different ion energies and find it to vary between R$_{\mathrm{ion}}$=18\,nm at 200\,eV  and R$_{\mathrm{ion}}$=22\,nm at 1\,keV.

Based on a linear extrapolation to the model of Verevkin et al.~\cite{Verevkin2002},
we can compare these radii to those expected for photons at the same energy R$_{\gamma}$($h\nu$ = E$_{\mathrm{ion}}$), which would amount to $R_{\gamma}$(0.2\,keV)=210\,nm and $R_{\gamma}$(1\,keV)=470\,nm, respectively. To quantify this difference between impacting matter and photons we introduce the \emph{conversion equivalent}
\begin{equation}\label{equ2}
\rho_c = R_{\mathrm{ion}}(E)/R_{\gamma}(E),
\end{equation}
which ranges between $\rho_c=5-10$\,\% in our case.

We may use the estimated hot-spot area to define a \emph{detection-equivalent photon energy} $E_d$, i.e. the energy a photon would need to create a hot-spot as large as observed for the ions. We find $E_d=1.8$\,eV for $E_{\mathrm{ion}}=200$\,eV and $E_d=2.6$\,eV for $E_{\mathrm{ion}}=1$\,keV.
We interpret this as an indication that by far the largest fraction ($>99$\,\%) of the incident ion energy is not transferred to the electronic system of the superconductor, at all.

\section{Role of the surface adsorbates}

Many reasons may be invoked for the incomplete transfer of kinetic energy:
While photons couple directly to the electronic system and only as a consequence of that to phonons as well, the incident ions are expected to interact with both systems at once. In the case of complex molecules, the impact may also redistribute energy between the translational and the internal degrees of freedom via inelastic deformations. One more dissipation channel is related to surface adsorbates that accumulate on the cold detector surface and act as a damping cushion, even at a base pressure of $10^{-8}$\,mbar.

In order to demonstrate the relevance of surface adsorbates we recorded the relative detection efficiency while varying the surface conditions of our detector by allowing helium to gradually condense on it:
With the ion gun switched off, we raised the residual Helium pressure in the chamber to $10^{-5}$\,mbar for about 60\,seconds. A few seconds later the pressure was restored to $10^{-7}$ mbar, and the ion gun was turned on again.

\begin{figure} [h]
  \centering
  \includegraphics[width=8.5 cm]{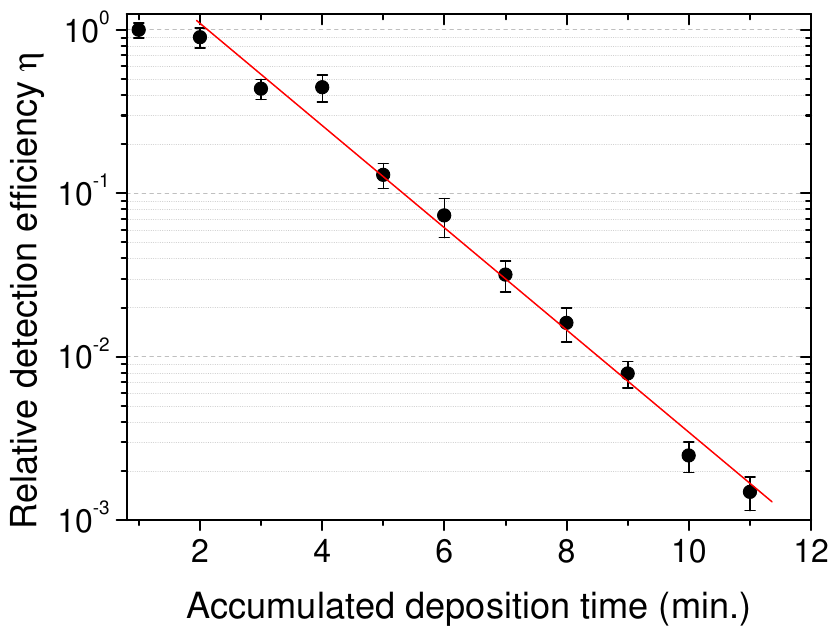}
  \caption{Relative detection efficiency for a 300\,eV helium ion beam as a function of the accumulated deposition time of neutral helium on the SSPD. The data show an exponential decrease of the detection efficiency with increasing surface coverage.}
 \label{Fig3}
\end{figure}

Figure~\ref{Fig3} shows the detection efficiency as a function of the integrated adsorbate accumulation time for He$^{+}$ ions at an energy of 300\,eV. In this setting, we find an exponential decay of $\eta$. With increasing surface coverage the efficiency drops by as much as a factor of one thousand. The causal relation between the decreasing detection efficiency and surface condensates is corroborated by the observation that we can largely restore the efficiency by locally heating the chip. Since the ohmic resistance of the nanowire is too small to dissipate enough power and to effectively heat the chip, we mounted a black absorber on the backside of the SSPD sapphire substrate and used a blue laser (447\,nm, 1000\,mW) for up to 5~minutes to partially evaporate the surface contaminations. This allowed us to regain a factor of one hundred in the detection efficiency.

\section{Prospective sensor applications}
In our discussions so far, we have focused on the detection efficiency as a function of the SSPD bias current (Fig.~2) or surface adsorbates (Fig.~3).
A quantitative characterization of $\eta$ is sufficient to understand and devise the SSPD as a detector for mass spectrometry.

\begin{figure} [h]
 \centering
  \includegraphics[width=8.5cm]{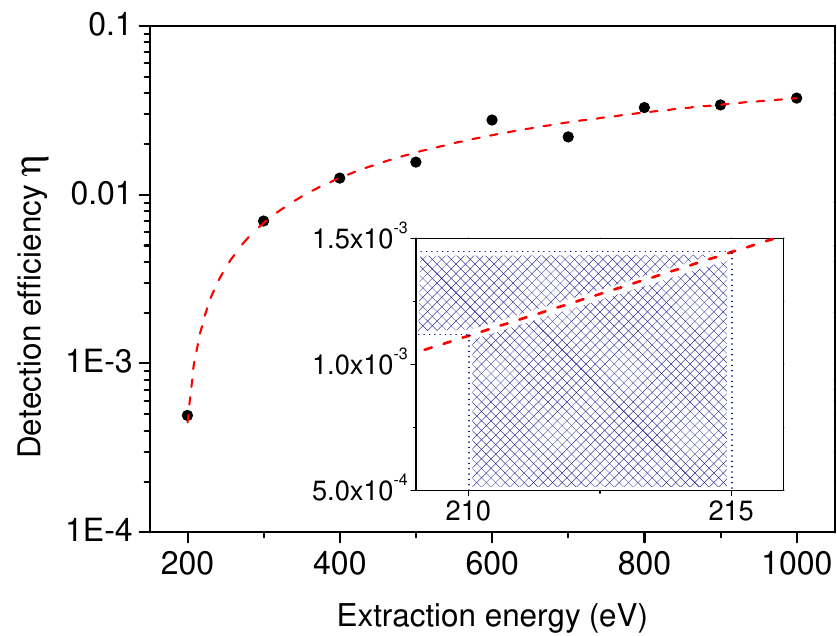}
 \caption{Detection efficiency as a function of the kinetic ion energy $E_{ion}$ at I$_{b}$/I$_{c}$=0.25. \textbf{inset}: an increase of few eV can be associated with a relative 50 $\%$ increase in the detection efficiency.}
  \label{Fig4}
\end{figure}

In the following we go one step further and propose a specific application in optical spectroscopy on molecules and nanoparticles. In order to elucidate this field, we refer to Figure \ref{Fig4} where we plot $\eta$ as a function of the impact energy for  I$_{b}$/I$_{c}$ = 0.25. The interesting information is revealed in the inset when we find that an energy variation by as little as 5\,eV may already lead to an increase in the detection efficiency of 50$\%$.

We note that in this particular experimental realization the detection efficiency is smaller than the one shown in Figure\,2.
At least two effects contribute to this observation: First, the surface coverage was not controlled in this experiment, as we were interested in relative efficiencies, only. Second, the critical current of each chip decreases with its extended exposure to energetic ions. The creation of a single local defect in the nanowire suffices to affect the detection efficiency. For the data shown in Figure\,4 the critical current I$_c$=10\,$\mu$A was half the value typically observed for new chips.
Although the sensitivity to surface contaminations and other detector specificities would render an absolute energy measurement difficult,
small energy differences may still be detected rather reliably.

We here suggest that it will be feasible to measure small variations in atomic or molecular energies.
If SSPDs can be made sensitive to the internal energy of nanoparticles -- which remains to be explored in future experiments -- the required energy modulation  could be imparted in a typical photon absorption spectroscopy experiment where the probability of the energy transfer is modulated by scanning the frequency of the incident light. Many large molecules and nanoparticles convert this excitation energy into vibrations and store it up to their impact on the detector.

One may also convert internal to kinetic energy before the impact: helium nanodroplets composed of $10^4-10^8$\,atoms \cite{Toennies2004,Pentlehner2009} may act as 380\,mK coolants for organic molecules and provide an ideal environment for spectroscopy. We here propose a new kind of depletion spectroscopy, where the absorbed photon triggers the evaporation of several thousand helium atoms. If the evaporation is isotropic the net velocity remains unchanged, while both mass and kinetic energy are reduced by several percent. This setting should open a new window to single-photon absorption spectroscopy.

\section{Conclusion}

Our experiments show that superconducting nanowires are capable of recording atomic and molecular ion signals at energies much lower than typically used in mass spectrometry. We have determined the absolute single particle detection efficiency and find that it is feasible to achieve a detection efficiency of 100\,\%, in our setting for ion energies below 1000\,eV. This saturation energy is still orders of magnitude larger than expected under the assumption that the ions couple dominantly to the electronic system of the superconductor.

We have identified several possible dissipation channels and discuss here for the first time the relevance of surface adsorbates.  Because of their transparency to visible and IR light, they went unnoticed in all earlier photon counting experiments. They might, however, also compromise optical detection experiments in the vacuum ultraviolet wavelength range, if the surface adsorbates contain a large fraction of oxygen (residual air).
For molecules even a thin damping layer can reduce the detection efficiency by several orders of magnitude. This observation is important for defining strategies, such as intermittent detector heating, to maintain proper surface conditions over extended periods of time.

The observed sensitivity of the SSPD to the particle energy may open a new avenue for optical spectroscopy on molecules, nanoparticles and even doped Helium nanodroplets which will be interesting for further studies in the future.

\section{Acknowledgments}
We acknowledge financial support through the Austrian FWF grant (Z149-N16,Wittgenstein) and MS acknowledges fruitful discussions with M. Ohkubo.

\section*{References}

\bibliography{references}

\end{document}